\documentclass[aps,twocolumn,groupedaddress]{revtex4-1}

\usepackage{hyperref}
\usepackage[usenames,dvipsnames]{color}
\usepackage{bm}

\usepackage{amssymb}
\usepackage{amsmath}
\usepackage{graphicx}
\usepackage{cancel}

\newcommand{\gnd}{\mathrm{4S_{1/2}}}
\newcommand{\fourP}{\mathrm{4P_{3/2} }}
\newcommand{\fiveP}{\mathrm{5P_{3/2}  }}
\newcommand{\TD}{T_\mathrm{D}}
\newcommand{\kB}{k_\mathrm{B}}

\begin{document}

\title{Low-temperature, high-density magneto-optical trapping of potassium \\ using the open 4S $\rightarrow$ 5P transition at 405\,nm}
\author{D.\ C.\ McKay}
\altaffiliation[Also at ]{the Department of Physics, University of Illinois, Urbana, Illinois 61801, USA}
\author{D.\ Jervis}
\author{D.\ J.\ Fine} 
\author{J.\ W.\ Simpson-Porco}
\author{G.\ J.\ A.\ Edge} 
\author{J.\ H.\ Thywissen}
\altaffiliation[Also at ]{the Canadian Institute for Advanced Research, Toronto, Ontario, M5G 1Z8 Canada}
\affiliation{Department of Physics, CQIQC, and Institute for Optical Sciences, University of Toronto, M5S1A7 Canada}

\date{\today}

\begin{abstract}
We report the laser cooling and trapping of neutral potassium on an open transition. Fermionic $^{40}$K is captured using a magneto-optical trap (MOT) on the closed $\mathrm{4S_{1/2}}  \rightarrow \mathrm{4P_{3/2}} $ transition at  767\,nm and then transferred, with unit efficiency, to a MOT on the open $\mathrm{4S_{1/2}}  \rightarrow \mathrm{5P_{3/2}  }$ transition at 405\,nm. Because the $\mathrm{5P_{3/2}  }$ state has a smaller line width than the $\mathrm{4P_{3/2}}$ state, the Doppler limit is reduced. We observe temperatures as low as 63(6)\,$\mu$K, the coldest potassium MOT reported to date. The density of trapped atoms also increases, due to reduced temperature and reduced expulsive light forces. We measure a two-body loss coefficient of $\beta = 2 \times 10^{-10}$\,cm$^3\,$s$^{-1}$, and estimate an upper bound of $8 \times 10^{-18}$\,cm$^2$ for the ionization cross section of the 5P state at 405\,nm. The combined temperature and density improvement in the 405\,nm MOT is a twenty-fold increase in phase space density over our 767\,nm MOT, showing enhanced pre-cooling for quantum gas experiments. A qualitatively similar enhancement is observed in a 405\,nm MOT of bosonic $^{41}$K. 
\end{abstract}

\maketitle

\section{Introduction \label{sec:intro}}

Magneto-optical trapping is a widely applied technique for creating cold, dense samples of neutral atoms. Ultracold gas experiments typically use a MOT for accumulation and pre-cooling, although laser cooling alone has been unable to create a quantum degenerate sample without a subsequent evaporative cooling step. Density in laser cooling is limited by repulsive radiation pressure due to re-absorption \cite{wieman:1990,*wieman:1992} and by collisional loss processes \cite{weiner:1999}. The temperature limit, in the simplest two-level theory, is the Doppler temperature, $\kB \TD=\hbar \Gamma/2$, where $\Gamma$ is the line width of the excited state \cite{wineland:1979,*letokhov:1981,*ashkin:1979}. Fortuitously, the multi-level structure of ground states can allow for sub-Doppler temperatures \cite{lett:1989,dalibard:1989}, but these effects are not seen in all atomic species. Particularly, sub-Doppler cooling is non-existent in $^{6}$Li \cite{shimizu:1991} and weak \cite{modugno:1999,*cataliotti:1998} or difficult to observe \cite{demarco:1999} in $^{40}$K. Since these are the only two stable fermionic alkali isotopes, and thus commonly used for the study of quantum degenerate Fermi gases, new cooling techniques would be beneficial.

Laser cooling on narrower lines can achieve lower temperatures, as has been demonstrated with earth alkaline atoms. In the case of $^{88}$Sr \cite{vogel:1999,*katori:1999}, the broad 30\,MHz cycling transition ($^1$S$_0\rightarrow ^1$P$_1$) at 461\,nm is used to capture atoms, followed by cooling on the narrow 7.5\,kHz forbidden transition ($^1$S$_0\rightarrow ^3$P$_0$) at 689\,nm.  This two-step process combines a large capture rate during the first stage with the low Doppler temperature of the second stage. Alkali atoms do not have forbidden transitions, however higher excited states do have smaller line widths.  For example, in potassium, the 5P state has a line width of 1.19\,MHz, roughly five times smaller than the 6.0\,MHz line width of the 4P state that has been used in all potassium laser cooling experiments to date. However, the 4S $\rightarrow$ 5P transition is not a cycling transition, and alternate decay channels could depolarize magnetic sublevels, increase the depumping effect of the cooling laser, and interrupt laser cooling mechanisms.

\begin{figure}[tb!]
\includegraphics[width=0.45\textwidth]{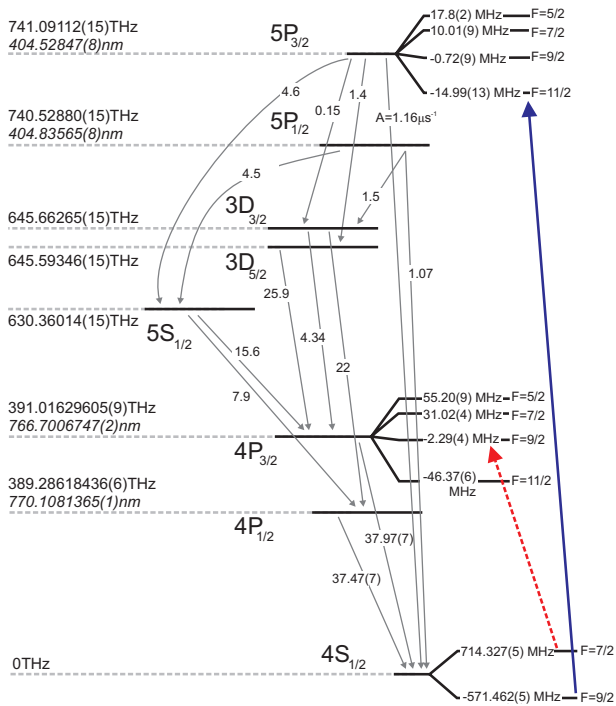}
\caption{ {\bf Level diagram} including all the possible decay channels of the 4S $\rightarrow$ 5P transition in $^{40}$K. Line strengths for each transition are from  \cite{sansonetti:2008} and listed as $A=1/\tau$ in units of $\mu s^{-1}$. Except for the 4P $\rightarrow$ 4S rates (from \cite{wang:1997}) the uncertainties in transition strengths are not shown explicitly. The uncertainty in the 5P $\rightarrow$ 4S rates are less than $8\%$ and all other rates are estimated to have uncertainties between 25\% and 50\%. The total measured lifetime from all decay channels is 134(2)ns for $\fiveP$ \cite{berends:1988} and 137.6(1.3)ns for $5P_{1/2}$ \cite{mills:2005}. Hyperfine splittings are from \cite{arimondo:1977} for $\gnd$ and $\fiveP$, and \cite{sansonetti:2008} for $\fourP$. Energy offsets from the 4S state are from \cite{sansonetti:2008} and presented in both frequency units (as $\omega/2 \pi$) and wavelength units ($\lambda$). The solid blue and dashed red arrows indicate the cooling and repump hyperfine transitions, respectively, used for laser cooling.
\label{fig:struc}}
\end{figure}

It has been shown that laser cooling is possible using open transitions, both with metastable helium \cite{Koelemeij:2003,*Tychkov:2004} and with lithium \cite{hulet:2011}. In the case of He*, 
a MOT on the 2S $\rightarrow$ 3P transition at 389\,nm was shown to have a lower two-body loss rate $\beta$, lower re-absorption rates, and a larger cooling force per recoil, resulting in increased density. No reduction in temperature was seen since the He* 2P and 3P excited states have the same lifetime. In the case of $^6$Li, a MOT on the narrower 2S $\rightarrow$ 3P line did have a reduced temperature, but not an increased density. Unlike lithium and metastable helium, potassium has a D-state decay channel that could perturb laser cooling more significantly (see Fig.~\ref{fig:struc}). In addition to reduced Doppler temperatures, higher-state transitions may provide other advantages: multi-photon decay channels allow for background-free detection, and the smaller transition wavelength reduces the diffraction limit of imaging and manipulation. A specific advantage of the potassium transition at 405\,nm is the emergence of inexpensive GaN diode lasers \cite{gustafsson:2000} in the 395--410\,nm range.

In this paper, we explore cooling and trapping on the ``blue'' $\gnd \rightarrow \fiveP$ transition of $^{40}$K at 405\,nm. Atoms are first accumulated in a standard MOT on the ``D2'' $\gnd \rightarrow \fourP$ transition at 767\,nm, for which the capture rate is high. The cloud is then transferred to a blue MOT, where we observe a temperature as low as 63(6)\,$\mu$K and a ten-fold increase in density. Although the capture velocity and beam size of the blue MOT are small, we observe almost perfect transfer between MOTs. We find that loss rates in the blue MOT from two-body loss and photoionization are higher than in the D2 MOT, but these processes play little role on the sub-100-ms timescale that we observe is necessary for equilibration.

Our paper is organized as follows. In Sec.~\ref{sec:theory} we discuss the state structure of potassium and calculate the effect of excitation on the $\gnd \rightarrow \fiveP$ transition using the steady-state Bloch equations. We estimate the cooling and trapping power with a semiclassical treatment. In Sec.~\ref{sec:setup} we describe the critical components of our experimental system, including generation of 405\,nm light with diode lasers and the optical configuration of the two-color MOT. Sec.~\ref{sec:mot} presents the primary results of the paper: the capture fraction, lifetime, temperature, and density of a blue potassium MOT. Further laser cooling results, prospects, and conclusions are discussed in Sec.~\ref{sec:conclusion}.

\begin{table}[t]
\caption{ {\bf Key excitation properties of the $\bm{\gnd \rightarrow \fiveP}$ transition}, with the $\gnd\rightarrow \fourP$ transition shown for comparison. Where relevant, the drive is assumed to be $\sigma^+$ polarized. Saturation intensity is defined as the drive strength at which 25\% of the population is the excited state. ``Steady-state polarization'' is the fraction of atoms in $F=9/2$ that are in the $m_F = +9/2$ sublevel, with a drive at $I_\mathrm{sat}$. ``Depumping probability'' is the percent decay from $|11/2,11/2\rangle$ to the $F=7/2$ manifold, through a multi-photon cascade. ``Doppler temperature'' is the steady-state temperature in the low-intensity limit, and includes open-transition effects for $\fiveP$. 
\label{table:transitionparams}
}
\begin{tabular}{|l|c|c|}
\multicolumn{1}{c}{ } & \multicolumn{1}{c}{ $\fourP$} & \multicolumn{1}{c}{ $\fiveP$ } \\ \hline 
Line width ($\Gamma / 2 \pi$) & 6.04(1)\,MHz & 1.19(2)\,MHz \\ \hline
Branching ratio & Cycling & 1\,/\,6.4(7) \\ \hline
Cross section ($\sigma$) & 0.28\,$\mu$m$^2$ & 0.010(1)\,$\mu$m$^2$ \\ \hline
Saturation intensity ($I_\mathrm{sat}$) & 1.752(3)\,mW/cm$^2$ & 23(2)\,mW/cm$^2$ \\ \hline
Steady-state polarization & 100\% & 61(1)\% \\ \hline
Depumping probability & 0\% & 18(2)\% \\ \hline
Doppler temperature & 145.0(3)\,$\mu$K & 23.5(7)\,$\mu$K \\ \hline
\end{tabular}
\end{table}

\section{4S$_{1/2}$ $\rightarrow$ 5P$_{3/2}$ Transition: Theory and Background  \label{sec:theory}}

Figure~\ref{fig:struc} shows a level diagram of potassium including all possible decay paths from the $\fiveP$ state. The line strength of the $\gnd \rightarrow \fiveP$ is $2 \pi \times 185$\,kHz, thirty times weaker than the $2 \pi \times 6.04$\,MHz D2 transition. However, the excited state is broadened to $\Gamma= 2 \pi \times 1.19$\,MHz by other decay channels: five out of six times the $\fiveP$ state decays via a three-photon cascade instead of emitting a single blue photon.

We calculate excitation properties and cooling rates of the 4S $\rightarrow$ 5P transition by finding the steady-state solution of the optical Bloch equation (OBE) that includes all relevant levels. Because the transition is not closed, continual excitation requires a repumping beam. As in the experiment, atoms are repumped on the $\gnd,F=7/2 \rightarrow \fourP,F=9/2$ transition, whose strength is chosen here to be powerful enough that it is not the limiting timescale. 

Table \ref{table:transitionparams} summarizes key steady-state properties of excitation to the 4P and 5P states. Unlike a cycling transition, polarized optical pumping on the open transition scrambles the $m_F$ states. However, the depolarization rate is not quite as bad as Fig.~\ref{fig:struc} might suggest: 61\% of atoms in the $F=9/2$ manifold are in the doubly-polarized $|9/2,9/2\rangle$ state. The resonant scattering cross-section is nearly thirty times smaller than the D2 transition due to a narrow line width and smaller wavelength. This could be an advantage for cooling trapped or dense clouds since optical density is reduced. However, for the same reasons, the saturation intensity is an order of magnitude higher, increasing optical power requirements.

\begin{figure}[t]
\includegraphics[width=0.45\textwidth]{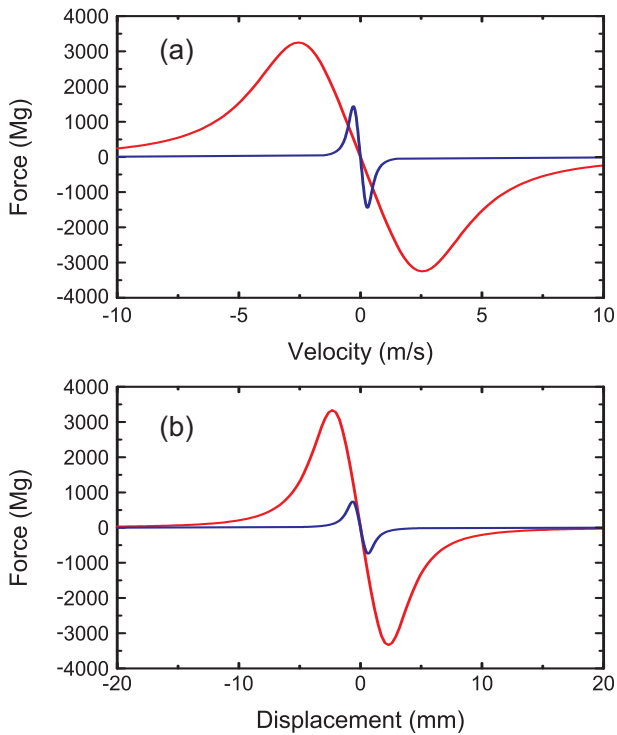}
\caption{ {\bf Blue cooling and trapping.} Calculated force (in units of $M g$, where $g$ is gravitational acceleration) versus {\bf (a)} velocity (at zero displacement) or {\bf (b)} position (at zero velocity) for a 10\,G/cm MOT at  767\,nm (red line) and 405\,nm (blue line). Both are calculated for counter-propagating beams at $-\Gamma/2$  detuning from resonance and with intensity per beam of $I_\mathrm{sat}/5$ where $I_\mathrm{sat}$ is indicated in Table \ref{table:transitionparams}. For the 405\,nm data there is a  767\,nm repump beam on resonance with the $\gnd,F=7/2 \rightarrow \fourP,F=9/2$ transition with 10\% of the 405\,nm beam intensity. Near zero velocity, the damping rate is $\gamma = 1.8 \times 10^4$\,s$^{-1}$ for the red and $\gamma = 6.9 \times 10^4$\,s$^{-1}$ for the blue. Near zero displacement, the spring constant corresponds to an undamped trap frequency $\omega_\mathrm{osc}/ 2 \pi = 730$\,Hz for the red and $\omega_\mathrm{osc}/ 2 \pi = 650$\,Hz for the blue.
\label{fig:force}}
\end{figure}

Laser cooling and trapping performance of the 4S $\rightarrow$ 5P transition is estimated by including either Doppler or Zeeman shifts in the OBE calculation. Recoil heating is assumed
to arise from momentum diffusion with an energy increase of $(\hbar k)^2/(2M)$ per scattering event, where $k=\omega/c$, $\omega$ is the transition frequency, $c$ is the speed of light, and $M$ is the mass of $^{40}$K. Heating in the multi-photon decay path is due to the scattering of three photons of various momenta. 

At comparable scaled detunings and saturations, Fig.~\ref{fig:force}(a) shows that the blue transition can achieve a higher cooling rate than the D2 transition. However, the maximum force and the capture range are greatly reduced. Figure~\ref{fig:force}(b) shows that the spring constant is comparable for the two transitions, but that trapping volume is larger for the red MOT. These calculations support the strategy of loading atoms using the 767\,nm transition and then transferring later to the 405\,nm MOT for additional cooling and compression. We find a steady-state temperature of 24\,$\mu$K for the blue transition, indeed smaller than the Doppler temperature of the D2 transition. In fact, this steady-state temperature in the low-intensity limit is slightly ($\sim$18\%) less than the $\hbar\Gamma/2 \kB$ expected in the two-level Doppler cooling, because the multi-photon cascade causes slower momentum diffusion than single-photon kicks with the same total energy.

Another difference between laser cooling with blue instead of infrared light is the possibility of photo-ionization. The ionization energy of potassium is 1049.56782(2)\,THz \cite{sansonetti:2008}, so the 5P state can be ionized by the 405\,nm MOT light, or by any photon with a wavelength less than $972$\,nm. By contrast, the 4P state requires photons with a wavelength less than $455$\,nm, which is not provided by the D2 MOT. The 405\,nm trap light may also ionize any state during the radiative cascade, and the 767\,nm repump can only ionize the 5P state. 

\begin{figure}[t!]
\includegraphics[width=0.5\textwidth]{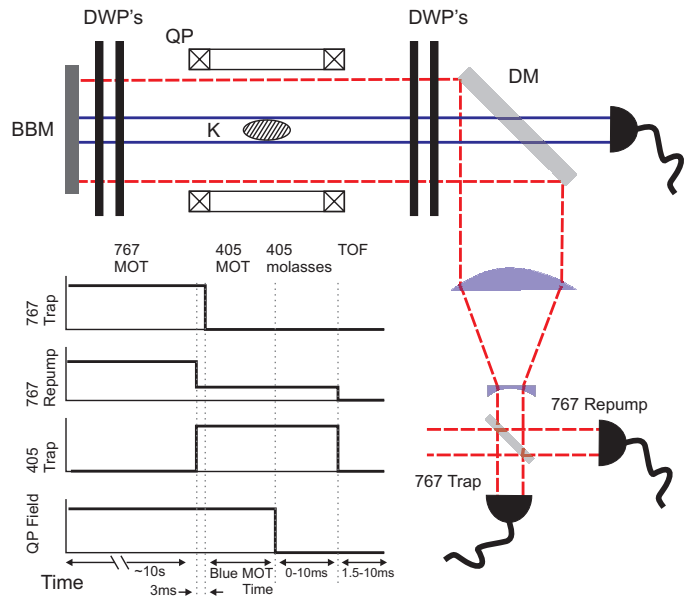}
\caption{ {\bf Experiment.} The optical configuration of a single arm of the magneto-optical trap includes 767\,nm trapping and repump beams (both shown with red dashed outlines) and a 405\,nm trapping beam (shown in solid blue). The two colors are mixed on a dichroic mirror (DM), manipulated with dichroic quarter-wave plates (DWP) that are wavelength-specific, and retro-reflected off a broad-band mirror (BBM). Only one MOT direction has been shown for clarity, but there is an identical beam path in three orthogonal directions. The magnetic quadrupole (QP) coils typically apply 10\,G/cm along their strong axis. The potassium (K) cloud is imaged in absorption with a probe beam (not shown).
{\bf Inset:} Timing diagram of laser beam intensities (top three lines) and magnetic field strength. After a 10\,s accumulation, the 767\,nm trap is extinguished, the 767\,nm repump intensity is reduced, and the 405\,nm trapping beam is turned on for a variable hold time. The molasses phase is only used in Sec.~\ref{sec:conclusion}. An absorption image is taken after a variable length time of flight (TOF).
\label{fig:motsetup}}
\end{figure}

\section{Laser and MOT Setup \label{sec:setup}}

The 4S $\rightarrow$ 5P transition is placed in a wavelength band for which 
GaN diode lasers have become readily available, due to recent commercial interest in optical storage media. These sources have been exploited to perform 4S $\rightarrow$ 5P spectroscopy of natural potassium \cite{gustafsson:2000,gustafsson:2000b,uetake:2003,halloran:2009,mills:2005} and enriched $^{40}$K \cite{behrle:2011}, but never laser cooling. 

Our 405\,nm master is a grating-stabilized diode laser with 10\,mW output. Half of this power is injected \cite{park:2003,*komori:2003} into a 100-mW diode laser (``slave'' laser).  At room temperature, and without injection, the slave diode spectrum is nearly 1\,nm wide and centered at 407\,nm.  The poor current and temperature tuning characteristics of the diode, approximately +0.02\,nm/mA and +0.05\,nm/$^{\circ}$C respectively, necessitate cooling the diode to -20\,$^{\circ}$C.  Without injection the Fabry-Perot spectrum of the slave is featureless, but strong peaks are evident when successfully injected. The output of the slave, typically run at 60\,mW, is passed through a single-mode optical fiber before use in the MOT (see Fig.~\ref{fig:motsetup}).

The remaining 5\,mW of master light is used for modulation transfer spectroscopy using a natural-abundance potassium cell heated to 140$^{\circ}$C. The elevated temperature is necessary because the Doppler cross section is two orders of magnitude smaller than the cross section for the D2 transition. We lock to the $\gnd,F=1 \rightarrow \fiveP$ transition in $^{39}$K (with unresolved excited states) that is the closest spectral feature to the $\gnd,F=9/2 \rightarrow \fiveP,F=11/2$ transition of $^{40}$K. The additional 0.5\,GHz shift is performed with acousto-optic modulators.

The D2 MOT uses retro-reflected beams along three axes, limited by the cell windows to be 4.4\,cm diameter in the horizontal axes and 3.8\,cm diameter in the vertical axis. A background potassium vapor is created from enriched dispensers (5\% $^{40}$K), and collected in the trap using the strong $\gnd \rightarrow \fourP$ transition. During this phase, there are two overlapping 767\,nm beams: a trap with 200\,mW total power detuned by $\Delta=-35$\,MHz from the $F=9/2 \rightarrow F'=11/2$ transition, and a repump with 150\,mW total power detuned by $-25$\,MHz from the $F=7/2 \rightarrow F'=9/2$ transition. After $10$\,s, the blue MOT is turned on, with up to 30\,mW total power. As shown in Fig.~\ref{fig:motsetup}, the smaller (10-mm diameter) blue beams are mixed with the infrared beams on dichroic mirrors. We use dichroic wave plates to control independently the polarization of the two colors. There is a short time (3\,ms) when both MOTs are on before the D2 trap light is turned off (see timing diagram of Fig.~\ref{fig:motsetup}), while  the D2 repump beam is left on at reduced intensity.  Blue laser cooling is applied typically for 30\,ms.

Laser-cooled atoms are characterized using absorption imaging on the 4S $\rightarrow$ 4P transition. Temperature is measured in time-of-flight, with a gaussian fit to the density distribution imaged at various free-flight times, 1.5--10\,ms after the magnetic field and laser beams have been extinguished.

\begin{figure}[t!]
\includegraphics[width=0.4\textwidth]{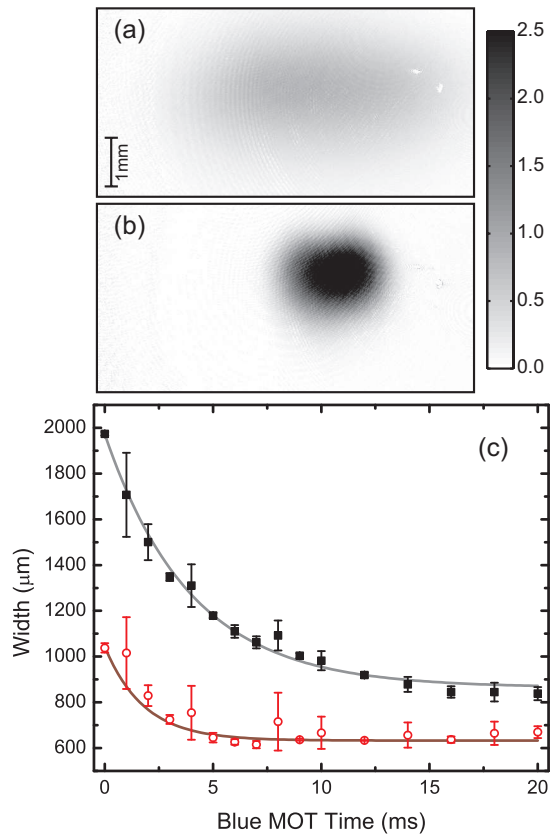}
\caption{ {\bf Trapped cloud measurements.} Typical absorption images of {\bf (a)} a 767\,nm MOT with detuning $\Delta=-35$\,MHz and intensity $I\approx 16$\,mW/cm$^2$; and {\bf (b)} a 405\,nm MOT, for $\Delta=-1$\,MHz, $I\approx90$\,mW/cm$^2$. In both images, pixel darkness denotes optical density, according to the scale bar at right. Images were taken shortly (1.5\,ms) after release from the trap, such that clouds are little changed from their in situ distributions. Atom number is $10^8$, but the cloud in (b) is clearly compressed. {\bf (c)} Cooling dynamics are observed in images similar to (b) but with various hold times in a blue MOT, here with $\Delta= -2$\,MHz and $I\approx50$\,mW/cm$^2$. The width is shown for both the vertical (open red circles) and horizontal (black squares) directions. Simple exponential fits are shown as solid lines, with a $1/e$ time of 1.8\,ms for the vertical width and 4\,ms for the horizontal width.
\label{fig:motabs}}
\end{figure}

\begin{figure}[tb!]
\includegraphics[width=0.4\textwidth]{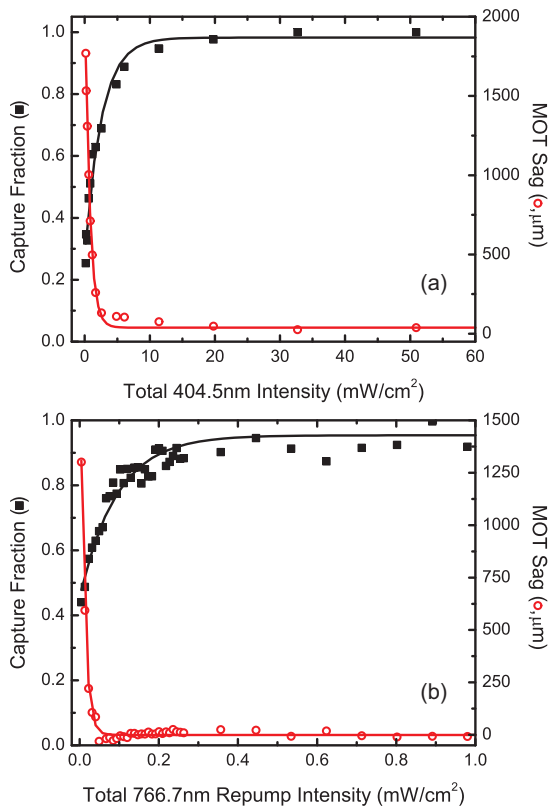}
\caption{ {\bf Intensity requirements.} A threshold for proper trap function is observed by measuring the capture fraction (black squares) and vertical displacement of the cloud from trap center (open red circles) as a function of the six-beam intensity of {\bf (a)} trap beams and {\bf(b)} repump beams. Lines are exponential fits to guide the eye. The blue-MOT time was $30$\,ms for (a) and $20$\,ms for (b), both at $-2$\,MHz detuning. In both sets of data the trap movement in the direction orthogonal to gravity was less than 100\,$\mu$m.
\label{fig:capture}}
\end{figure}

\section{4S $\rightarrow$ 5P MOT of $^{40}$K \label{sec:mot}}

We observe that atoms are held by the blue MOT. A typical absorption image of the D2 MOT and the blue MOT are shown in Fig.~\ref{fig:motabs}. Across a broad range of parameters, compression occurs, which is a first indication of cooling (see below for further discussion). 
Fig.~\ref{fig:motabs}(c) shows that cloud sizes in the blue MOT equilibrate with a $1/e$ relaxation time of 2--4\,ms. In a damped oscillator model of atomic motion in a MOT, where $\gamma \gg \omega_\mathrm{osc}$, the atomic position damps with a time $\gamma /\omega_\mathrm{osc}^2$, which is consistent with what we measure --- 4\,ms using the typical force curves of Fig.~\ref{fig:force}(b). Atomic velocities would damp in a much faster time $\gamma^{-1} \sim 15\,\mu$s, so after the spatial compression shown in Fig.~\ref{fig:motabs}(c), blue laser cooling is complete.

\begin{figure}[tb!]
\includegraphics[width=0.45\textwidth]{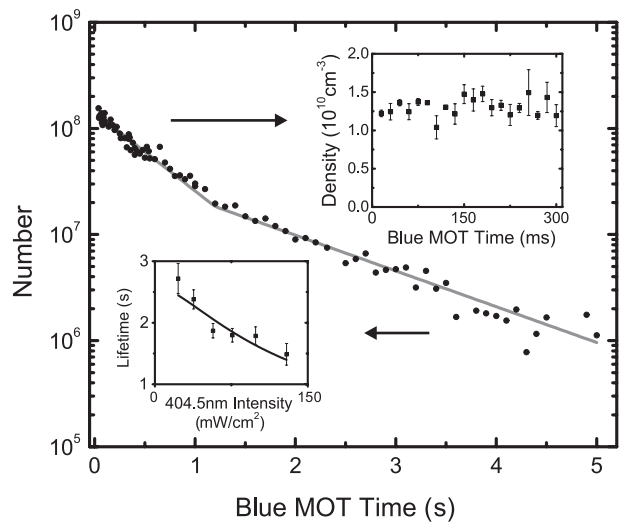}
\caption{{\bf Lifetime.} Atom number is shown versus hold time in the blue MOT. The data are fit to exponential decays in different regions as a guide to the eye. {\bf Top inset:} Peak density for short hold times is approximately constant. {\bf Bottom inset:} The fit one-body lifetime at long hold times decreases at higher 405\,nm trap intensity. 
\label{fig:lifetime}}
\end{figure}

\begin{figure}[t!]
\includegraphics[width=0.45\textwidth]{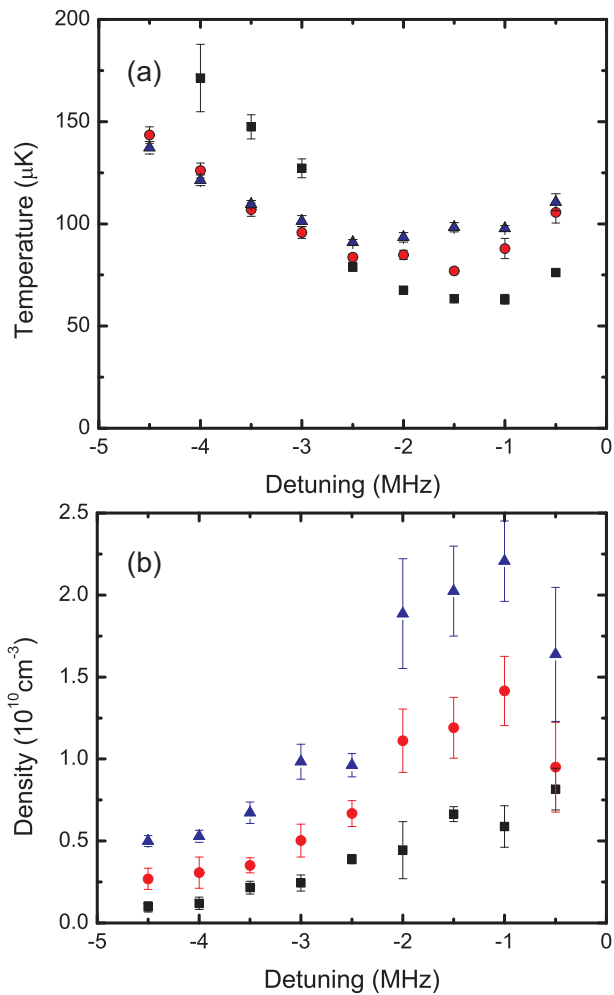}
\caption{  {\bf Performance of a magneto-optical trap at 405\,nm.} Both {\bf (a)} temperature and {\bf (b)} peak density are shown versus detuning from resonance, for several magnetic quadrupole gradients: 5\,G/cm (black squares), 7.5\,G/cm (red circles), and 10\,G/cm (blue triangles). Temperature is shown along a weak axis of the quadrupole; temperatures along the strong axis (and along gravity) were 1.82(15) times higher.
\label{fig:mot_temp_density}}
\end{figure}

\begin{figure}[t!]
\includegraphics[width=0.45\textwidth]{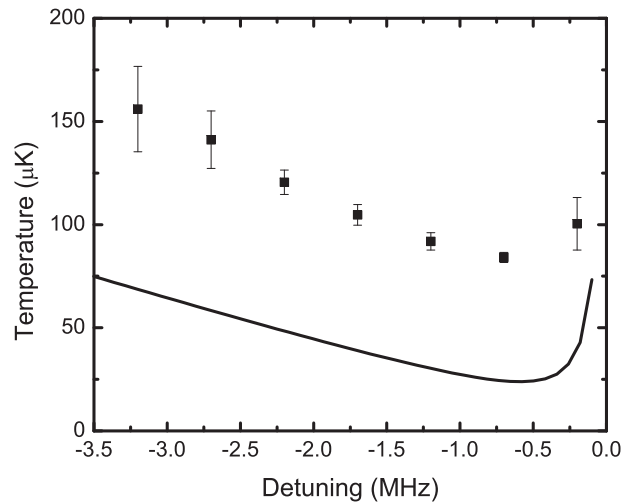}
\caption{{\bf Molasses cooling.} Blue laser cooling is continued for 10\,ms after the magnetic quadrupole field is extinguished, followed by a TOF temperature measurement. The horizontal temperature is shown versus detuning; vertical (along gravity) temperatures are 1.96(11) times higher. In addition to the statistical error bars shown, there is a systematic 0.5\,MHz uncertainty in the detuning. The solid line is an OBE calculation, but is not a quantitative prediction: the approach is constrained to intensities well below saturation (here, $I_\mathrm{sat}/50$ per beam), whereas the data was taken with roughly $I_\mathrm{sat}/2$ per beam.
\label{fig:temp}}
\end{figure}

Figure~\ref{fig:capture} shows the capture fraction and gravitational sag as a function of blue trap intensity and D2 repump intensity. In both cases, providing the saturation intensity (see Table \ref{table:transitionparams}) is sufficient for complete transfer into a strong trap.%
\footnote{For low intensities, the measured sag values are roughly ten times larger than the sag one would expect from our simple model. Also note that the displacement switches sign for intermediate repump intensity, which would not be the case if the spring constant were monotonic in intensity. It may be that optical pumping is not correctly functioning in a MOT at intensities far below saturation.}
From exponential fits, the transfer efficiency is 90\% for 5.5\,mW/cm$^2$ of blue trap intensity and for 0.21\,mW/cm$^2$ of D2 repump intensity. The 26:1 ratio between these follows approximately the ratio of saturation intensities of each transition. 

Since the capture rate of a MOT from background vapor is proportional to $\Gamma^4 d^4$ for small beam sizes $d$, one would expect a $10^6$ reduction in steady-state number in going from the 767\,nm to the 405\,nm trap. Accordingly, a measurement of atom number versus time (Fig.~\ref{fig:lifetime}) shows no sign of saturation after a hundred-fold reduction in number. We can then use such a measurement as a clean measure of loss processes, ignoring additional capture.

We also note that since we are not loading atoms from vapor into the blue MOT, we can choose small beam sizes, and reduce the 405\,nm power requirement. In our configuration, the total power required at saturation intensity is roughly 2.5\,mW, which our laser system can easily provide.

The loss data in Fig.~\ref{fig:lifetime} fits well to two exponential time scales, indicative of two-body losses at high density, and one-body losses (collisions with background atoms plus photoionization) at low density. The loss rate from a MOT is \cite{weiner:1999}
\begin{equation}
\frac{dN}{dt} = -\frac{N}{\tau} - \beta \int n^2 dx,
\end{equation}
where $\tau$ is the one-body lifetime, $\beta$ is the two-body loss coefficient, and $n$ is the atomic number density. To find $\tau$, we fit the data at long hold times ($t>2$\,s, $N<5\times 10^6$) where the atom number is strongly reduced and two-body losses can be neglected. The lifetime versus 405\,nm intensity is show as the bottom inset to Fig.~\ref{fig:lifetime}. The decrease in lifetime versus intensity could be evidence of ionization losses. The line is a fit to a model that includes the intensity dependence of both the ionization rate and the excited state fraction. A separate study of MOT fluorescence versus laser power, comparing D2 and blue MOTs, was used to calibrate excited state fraction versus intensity. Together, we find an ionization cross section of $8\times10^{-18}$cm$^2$ for the $\fiveP$ state at 405\,nm. We report this value as an upper bound since the ionizing beams are also the trap beams and we cannot rule out other trap loss processes
\footnote{From similar measurements of lifetime versus repump intensity, we constrain the effect of repump ionization to be negligible. One would guess a 15\,s ionization time for $I_{767}=1$\,mW/cm$^2$ using the rubidium ionization cross section of $\sigma=1.5 \times 10^{-17}$\,cm$^2$ measured at 690\,nm by \cite{ambartzumian:1976}}.

There are no measured cross sections for the ionization of the 5P states of potassium, only for the equivalent state, 6P, of rubidium \cite{anderlini:2004,courtade:2004,ambartzumian:1976}. These rates may be comparable since the ionization cross section from the potassium 4P states \cite{petrov:2000,amina:2008} is within a factor of three of the 5P cross section for rubidium \cite{dinneen:1992}. Indeed, our upper bound is similar in magnitude to the $2\times10^{-18}$cm$^2$ cross section for the 6P state of rubidium from \cite{ambartzumian:1976} measured at $\lambda=350$\,nm.

To measure the two-body loss rate, we assume the MOT density distribution is Gaussian, which is supported by absorption images of the cloud. From the top inset in Fig.~\ref{fig:lifetime} we see that the peak density $n_0$ at short times is roughly constant, so the number decays as
\begin{equation}
N(t) = N_0 \exp{\left[-\left(\frac{1}{\tau}+\frac{n_0 \beta}{\sqrt{8}}\right)t\right]}.
\end{equation}
We then fit the data at short hold times ($t<300$\,ms) to an exponential decay and use $\tau$ (from our previous measurement) and $n_0$ to find $\beta$. At  $\Delta=-2$\,MHz, we find $\beta=2.0(1)\times 10^{-10}$cm$^{3}$/s for $I\approx75$\,mW/cm$^2$, and $\beta=1.4(1)\times 10^{-10}$\,cm$^{3}$/s for $I\approx20$\,mW/cm$^2$. These are relatively high two-body loss rates compared to alkali MOTs on the D2 transition, which are typically in the range of $10^{-12}$ to $10^{-11}$\,cm$^{3}$/s \cite{weiner:1999}. Values for $^{40}$K  767\,nm MOTs vary from $6\times 10^{-13}$ \cite{ridinger:2011} to $3\times 10^{-11}$cm$^{3}$/s \cite{modugno:2003}. Therefore, overall loss rates in the 405\,nm MOT are larger than the  767\,nm MOT, but for the short timescales (tens of milliseconds) required to decrease temperatures and increase density, these loss rates are still negligible.

A primary motivation for cooling on a narrow transition is to achieve a lower asymptotic temperature. Figure~\ref{fig:mot_temp_density} shows how the temperature and peak density of the blue MOT change with detuning, at various quadrupole gradients. A minimum temperature of 63(6)\,$\mu$K is seen for a 5\,G/cm magnetic quadrupole gradient. Operating the MOT with higher gradients increases both the temperature and density.  In all cases,  the minimum temperature and maximum density are found near $\Delta/\Gamma \approx -1$.

We attribute the increase in density to three factors. First, the temperature in the MOT is lower. Second, expulsive forces due to re-absorption are reduced, because the same spring constant is achieved with a scattering rate that is six times smaller. The open transition may further reduce re-absorption by pumping to the $F=7/2$ ground states. Third, the optical density seen by both the incident and the scattered blue photons is reduced by a factor of nearly thirty.

In comparison, our D2 MOT density is typically $n_0 = 2\times10^9$\,cm$^{-3}$ and temperature is typically $T=180\,\mu$K. Since phase space density scales as $n_0 T^{-3/2}$, the combined $75\,\mu$K and $n_0 = 1.2\times10^{10}$\,cm$^{-3}$ (achieved with a blue MOT at 7.5\,G/cm and -1.5\,MHz detuning) realizes a twenty-fold enhancement in phase space density.

\section{Prospects and Conclusions \label{sec:conclusion}}

Since the lowest observed MOT temperatures occur at low gradients, one might expect continued improvement during an optical molasses phase, in which magnetic fields have been extinguished (see Fig.~\ref{fig:motsetup} inset). Figure~\ref{fig:temp} shows that minimum temperature, roughly $85\, \mu$K, is comparable to what is observed in the MOT, although optimized at a lower detuning, $\Delta / \Gamma = -0.6(2)$. These temperatures are still more than three times the Doppler temperature calculated in Sec.~\ref{sec:theory}, which may be due to a low cooling rate as compared to the short molasses time available. Indeed we observe that the temperature asymmetry between the vertical and horizontal axes in the MOT persists during the molasses phase, when no gradient asymmetry remains. Even the asymptotic temperature might not be expected to reach the Doppler limit: while the calculation in Sec.~\ref{sec:theory} includes multi-level effects, it neglects three-dimensional effects, reabsorption, and heating due the intensity fluctuations. The latter effect has been shown by Chaneli\`ere {\it et al.} to limit one-dimensional Doppler cooling to several times $\TD$ in Sr, when Sisyphus and other sub-Doppler mechanisms are absent \cite{chaneliere:2005}. In the case of 4S $\rightarrow$ 5P cooling in $^{40}$K, polarization gradient cooling may be interrupted by the optical pumping and depolarization effects of the three-photon cascade.

We have also observed laser cooling and trapping for bosonic $^{41}$K on the $\gnd \rightarrow \fiveP$ transition. A similar order-of-magnitude increase in MOT density was apparent, although further study is warranted. With any isotope of potassium, additional improvements might be possible with more sophisticated timing sequences of detuning and/or intensity of cooling light, which has been fruitful in narrow-line MOTs and in sub-Doppler cooling of bosonic potassium \cite{inguscio:2011}.

In sum, we have observed laser cooling and trapping of neutral potassium on the open $\gnd\rightarrow \fiveP$ transition. Nearly complete transfer from a D2 MOT is observed, although the blue MOT is ineffective at capturing directly from the vapor. We observe temperatures as low as 63(6)\,$\mu$K along one axis, roughly half of the Doppler limit on the D2 transition. Unlike sub-Doppler cooling observed in $^{40}$K during a molasses phase \cite{modugno:1999}, our approach reduces temperature while the confining MOT quadrupole field is still present. Density is enhanced ten-fold in typical conditions, due to lower temperature and reduced re-absorption effects. Cooling and compression together increase the phase space density by more than an order of magnitude, demonstrating improved pre-cooling for quantum degenerate gas experiments. The reduced wavelength of the $\fourP \rightarrow \fiveP$ transition may enable improved imaging or addressing resolution ($\lambda/2=202$\,nm), for instance in the context of  strongly correlated lattices where the spacing is typically 500\,nm \cite{bakr:2009,*sherson:2010}.
More generally, since the potassium cascade structure is shared by sodium, rubidium, cesium, and francium, our work along with Ref.~\cite{hulet:2011} suggests that laser cooling and trapping using the $n$S $\rightarrow$ $(n+1)$P transition will be successful for all alkali atoms. \\

\begin{acknowledgments}
We would like to thank K.\ Pilch, M.\ Sch\"oll, and F.\ Stubenrauch for their early contributions, and thank B.\ DeMarco, R.\ Hulet, P.\ Duarte, A.\ Steinberg, M.\ Greiner, and S.\ Trotzky for discussions. This work was supported by the DARPE OLE program, ARO, AFOSR, CIfAR, CFI, and NSERC.
\end{acknowledgments}

\bibliography{blue_motV8}

\end{document}